# Proximity-Induced Superconductivity with Subgap Anomaly in Type II Weyl Semi-Metal WTe$_2$

Qiao Li[†], Chaocheng He[†], Yaojia Wang[†], Erfu Liu[†], Miao Wang[†], Yu Wang[†], Junwen Zeng[†], Zecheng Ma[†], Tianjun Cao[†], Changjiang Yi[‡], Naizhou Wang[§,∥], Kenji. Watanabe[⊥], Takashi. Taniguchi[⊥], Lubing Shao[†], Youguo Shi[‡], Xianhui Chen[§,∥], Shi-Jun Liang[*,†], Qiang-Hua Wang[*,†], Feng Miao[*,†]

[†] National Laboratory of Solid State Microstructures, School of Physics, Collaborative Innovation Center of Advanced Microstructures, Nanjing University, Nanjing 210093, China.

[‡] Institute of Physics, Chinese Academy of Sciences, Beijing 100190, China.

[§] Hefei National Laboratory for Physical Science at Microscale and Department of Physics, University of Science and Technology of China, Hefei, Anhui 230026, China.

[∥] Key Laboratory of Strongly Coupled Quantum Matter Physics, University of Science and Technology of China, Hefei, Anhui 230026, China.

[⊥] National Institute for Materials Science, 1-1 Namiki Tsukuba, Ibaraki 305-0044, Japan.

Corresponding Authors
*Tel: +86-025-83621497. Fax: +86-025-83621497. E-mail: miao@nju.edu.cn (F. M.); qhwang@nju.edu.cn (Q. H. W.); sjliang@nju.edu.cn (S. J. L.).

**ABSTRACT:** Due to the non-trivial topological band structure in type-II Weyl semimetal Tungsten ditelluride ($WTe_2$), unconventional properties may emerge in its superconducting phase. While realizing intrinsic superconductivity has been challenging in the type-II Weyl semimetal $WTe_2$, proximity effect may open an avenue for the realization of superconductivity. Here, we report the observation of proximity-induced superconductivity with a long coherence length along c axis in $WTe_2$ thin flakes based on a $WTe_2$/$NbSe_2$ van der Waals heterostructure. Interestingly, we also observe anomalous oscillations of the differential resistance during the transition from superconducting to normal state. Theoretical calculations show excellent agreement with experimental results, revealing that such a sub-gap anomaly is the intrinsic property of $WTe_2$ in superconducting state induced by the proximity effect. Our findings enrich the understanding of superconducting phase of type-II Weyl semimetals, and pave the way for their future applications in topological quantum computing.

Weyl semimetal, hosting low-energy quasiparticle excitations - Weyl fermions[1], is a new phase of matter with non-trivial topological band structure[2-4]. With linear band crossings around pairs of Weyl points (also called Weyl cones) and an open Fermi arc in surface state[2, 5-7], Weyl semimetals exhibit many unique properties, such as chiral anomaly[8-12] and anomalous Hall effect[13]. In addition, the multiple pairs of Weyl cones in the topological band structure may facilitate the formation of intra-cone pairing state with finite momentum and give rise to topological superconductivity[14-26] in Weyl semimetals when entering superconducting phase, which holds promise for fault-tolerant topological quantum computation. Thus, the potential realization of topological superconductivity in Weyl semimetals has sparked intense research interest.

Comparing to Type-I Weyl semimetals with a closed point-like Fermi surface, type-II Weyl semimetals possess unique features such as titled Weyl cone-like energy spectrum and paired Weyl nodes formed at the boundary of electron and hole pockets[27-31], which lead to planar orientation-dependent chiral anomaly[32]. More interestingly, a type-II Weyl semimetal is expected to exhibit distinctive superconducting properties from type-I Weyl semimetals due to the tilted Weyl cone energy spectrum[33]. As a typical type-II Weyl semimetal[27], Tungsten ditelluride ($WTe_2$) possesses distorted structure of $T_d$ phase (space group $Pnm2_1$) which breaks the inversion symmetry and accommodates the existence of type-II Weyl points in film and topological edge state in monolayer[34-39]. With the unique topological bands, $WTe_2$ is expected to offer a platform to study topological superconductivity. However, the intrinsic superconductivity in exfoliated $WTe_2$ thin film has not yet been observed[40].

In this work, we report proximity effect induced superconductivity in thin $WTe_2$ flakes. The observed superconductivity is found to give rise to a long superconducting coherence length (above 30 nm) in $WTe_2$ along its *c* axis. In addition, we observe anomalous oscillations in the differential resistance spectra as the bias current drives the system from superconducting to normal state. Moreover, our theoretical calculations show that the anomaly could be associated with the sub-gap features in the density of states (DOS) spectra in superconducting $WTe_2$. Our work enriches the understanding of superconducting phase in type-II Weyl semimetals and paves the way towards further exploration of topological superconducting states.

To realize proximity-induced superconductivity, a high-quality interface between thin $WTe_2$ flake and superconductor is desirable. To that end, we chose layered niobium diselenide ($NbSe_2$) as the superconductor (with $T_c$ ~ 7.2 K and $\Delta$ ~ 1.2 meV[41]) and fabricated $WTe_2$/$NbSe_2$ van der Waals heterostructure via dry-transfer method (see Methods for details)[42]. To avoid device degradation, both $NbSe_2$ and $WTe_2$ samples were exfoliated and transferred onto pre-prepared electrodes in glovebox. The whole heterostructure was further covered by a hexagonal Boron Nitride (h-BN) flake to avoid undesirable oxidation (as shown in Figure 1b). In the fabrication process, we selected the thin $WTe_2$ flakes with thickness of around 10 nm, to make sure that the flake was thick enough to exhibit band structures of type-II Weyl semimetal and thin enough to facilitate superconducting proximity effect. Besides, thin $NbSe_2$ flake (around 20 nm) with smaller area than $WTe_2$ sample was chosen to avoid $NbSe_2$ contacting with electrodes.

We first examined the low temperature transport properties of thin $WTe_2$ flakes by using a four-probe lock-in setup (with an AC current of 1 µA) (see Figure 1b). As shown in Figure 1c, we observe that the resistance of a typical device sharply drops to zero at a $T_c$ (critical temperature) of 6.4 K at zero magnetic field, thus the superconducting phase in $WTe_2$ is achieved, which has been verified in a different device setup (see the part I in Supporting Information). As the applied perpendicular magnetic field increases, the critical temperature $T_c$ reduces until the superconductivity is completely suppressed at around 5 T. The magnetoresistance curves at four different temperatures (T=1.57, 3, 5 and 7 K) are shown in Figure 1d. As temperature arises, the critical magnetic field becomes smaller. When the temperature increases to 7 K, the superconductivity disappears. A feature of weak anti-localization effect in the nonlinear magnetoresistance curve was then observed, similar to what has been reported in the normal resistance state of $WTe_2$[43]. Note that the proximity effect induced superconductivity in this work is related to the intrinsic band structure of $WTe_2$ as a type-II Weyl semimetal, different from high pressure-induced superconductivity in the 1T' phase WTe$_2$[44-47] and superconductivity in the epitaxial WTe$_2$ thin film on a sapphire substrate[48].

The achievement of superconductivity in the 10 nm-thick sample implies a long superconducting coherence length along *c* axis in WTe$_2$. Based on the measured mobility along *a* axis $\mu \sim 2100$ cm$^2$/V$^{-1}$s$^{-1}$(see Fig.S6 in Supporting Information), the mobility along *c* axis is estimated to be around 210 cm$^2$/V$^{-1}$s$^{-1}$ since it is one order magnitude smaller than that along *a* axis [49]. Combining the Fermi momentum $k_F \sim 1.2$ nm$^{-1}$ along *c* axis[49], the mean free path ($l_N$) along *c* axis in WTe$_2$ is calculated to be around 18 nm (according to $\mu = el/\hbar k_F$), which is larger than the superconducting coherence length ($\xi_S$) (2.3 nm) along *c* axis of NbSe$_2$[50]. In the "clean limit" (i.e. $\xi_S \ll l_N$)[51], the superconducting coherence length in WTe$_2$ $\xi_N$ can be calculated by $\xi_N(T) = \hbar v_N / 2\pi k_B T$, where $\hbar$ and $k_B$ represent Planck constant and Boltzmann constant respectively, $v_N$ is the Fermi velocity ($1.7 \times 10^5 \sim 3.1 \times 10^5$ m/s[52, 53]) and $T$ is the temperature (6.4 K). $\xi_N$ is estimated to be above 30 nm, which results from the intrinsic band structure of superconducting WTe$_2$ as a type-II Weyl semimetal, and will be discussed later.

To further investigate the proximity-induced superconducting properties in WTe$_2$, we performed differential resistance measurements (*dV/dI* vs. *V*) at various temperatures, with typical results shown in Figure 2a. Interestingly, we observe an anomalous feature in the differential resistance spectra when the temperature is below T$_c$. In the differential resistance spectrum, besides two remarkable symmetric peaks appearing as expected at around 280 µV, several small peaks emerge at lower voltage biases. Similar differential resistance spectrum has been observed in other samples (see Fig.S7 in Supporting Information). Different from that *dV/dI* vs. *V$_{ds}$* curve represents superconducting gap in conventional N/S junction, here, by applying an AC excitation current of 1 nA added on top of a variable DC bias current, *dV/dI* spectrum in this paper implies how WTe$_2$ resistance changes as the bias current drives WTe$_2$ from superconducting state to normal state. As temperature rises, these small peaks shrink in both amplitude and bias gradually, and eventually vanish together with two main peaks after reaching $T_c$ (6.4 K). The *I-V* characteristics at different temperatures can be obtained after integrating *dV/dI-I* curves, with results shown in Figure 2b. With increasing temperature, the superconductivity is completely suppressed at $T_c$~6.4 K, which agrees well with the differential resistance spectra. At 1.6 K, zero-resistance state occurs at a critical current $I_c$~125 µA ($I_c$ is defined as the current at which the voltage reduces to zero), and superconductivity in WTe$_2$ is suppressed at $I$~220 µA. Moreover, the measured $I_c$ (around 125 µA at $R_N$~2.5 Ω, with $R_N$ defined as the sample resistance above $T_c$) of the device is one order of magnitude smaller than the $I_c$ (1.8 mA at $R_N$~1 Ω) of NbSe$_2$ and this has been observed in other measured devices (see Table.S1 and Fig.S5 in the Supporting Information). Furthermore, the nonlinear variation of *I-V* curve between superconducting and normal state is slowly-varying, which is quite different from the abrupt change in NbSe$_2$.

To shed light on the observed anomaly during the transition from superconducting to normal state, we further investigated the magnetic field dependence of the differential resistance spectra and *I-V* characteristics at T=1.6 K (See Figure 2c and 2d). Exceptionally, we find that the two main peaks become broad as B increases within small magnetic field range (lower than 0.1 T), in contrast to the temperature-induced behavior. Furthermore, we observe that all sub-features in spectra vanish at 0.3 T and that only two broad symmetric peaks are retained. Similarly, *I-V* characteristics (in Figure 2d) at various magnetic fields can be obtained by integrating the *dV/dI-I* curves. At lower magnetic fields, the *I-V* curves also show two current points indicating critical current and transition to normal state. At higher magnetic fields, the linear parts indicating the normal state in *I-V* curves are obviously separated by various magnetic fields, consistent with the magnetic field dependence of normal resistance of WTe$_2$.

We calculated the DOS of type-II Weyl semimetal and s-wave superconductor in the WTe$_2$/NbSe$_2$ heterostructure to understand the anomalous experimental features of *dV/dI* spectra with results shown in the Figure 3a. WTe$_2$ and NbSe$_2$, respectively, are simulated by an effective two-orbital model with spin-orbital coupling and a simple one-orbital model with s-wave pairing in the inset of Figure 3a. (The calculation methods can be seen in the Supporting Information). We assume that there is no pairing potential in $T_d$-phase WTe$_2$, and that all of its superconducting behaviors are induced by coupling to the NbSe$_2$ superconductor. The DOS in superconducting NbSe$_2$ (blue line) is U-shaped, which is a feature of s-wave superconductor. However, the DOS (red line) in superconducting WTe$_2$ develops two coherence peaks at E =$\pm 0.25\Delta$, roughly one fourth of the gap in NbSe$_2$, which is identified as the induced superconducting gap of WTe$_2$. To have a closer view, we enlarge and replot the DOS of superconducting WTe$_2$ in Figure 3b. Inside the induced gap, we observe spiky sub-gap features consistent with the observed sub-features in the *dV/dI* vs *V* spectra in the Figure 2b and 2d. Besides, our theoretical calculations show that the gap features persist beyond ten layers in the WTe$_2$ while similar features already vanish in a type-I Weyl semimetal beyond the second layer (See Fig.S11 in the Supporting Information). We speculate that this may be related to the fact that the electron and hole pockets in WTe$_2$ provide larger DOS at the Fermi energy for electrons to pair up in proximity to the bulk superconductor, leading to a longer coherence length along *c* axis as aforementioned.

Based on the calculated results, we can understand the sub-gap anomaly observed in the differential resistance spectra in the following way. The total current in the sample is composed of the supercurrent and normal current when the bias current goes beyond the critical current. The dissipative normal current is carried by quasi-particles, which are excited by breaking the Cooper pairs self-consistently via the voltage drop induced by the normal current (as well as the Doppler shift effect in the presence of the current). The induced gap on the electron and hole Fermi pockets of WTe$_2$ is non-uniform, as evidenced by the peaks in DOS in Figure 3b (red line). The Cooper pair-breaking is a multiple-step process and starts from the smaller gap, instead of in an abrupt way as in conventional superconductors. The de-pairing of Cooper pairs at voltages corresponding to the DOS peaks excites a relatively larger number of quasiparticles, and the dissipative component of the total driving current will be enhanced, leading to an enhancement of the differential resistance. We emphasize that the voltage is measured under a bias current in our case, which is different from the usual case in which current is measured in response to a bias voltage.

To further prove that the unusual differential resistance spectra of superconducting WTe$_2$ thin flake root in the unique DOS of WTe$_2$, we carried out both theoretical and experimental studies on graphene/NbSe$_2$ device with 10 nm thick graphene flake (the device structure is shown in the inset of Figure 4b). We first calculated the DOS of the graphene and NbSe$_2$, with results shown in Figure 4a. The DOS of graphene resembles that of superconducting NbSe$_2$ but differs from that of superconducting WTe$_2$. We then measured the resistance of the graphene thin flake in a typical graphene/NbSe$_2$ device at various temperatures. As shown in Figure 4b, with decreasing temperature, superconductivity can be also achieved in graphene at a transition temperature T$_c$ of around 6.3 K, close to that of NbSe$_2$. In contrast to the magnetoresistance results in WTe$_2$, the normal resistance of graphene thin flake remains almost constant over the applied magnetic field range, which is consistent with the suppression of weak localization in graphene thin flake[54]. We further measured the differential resistance spectra at various temperatures and magnetic fields (see Figure 4c and Figure 4d). As expected, only two neat peaks are present in the scanning current range, without any sub-peaks emerging. When increasing temperature or magnetic field, the two peaks behave similarly to that of WTe$_2$.

Compared with the properties of superconducting graphene, the DOS of WTe$_2$ in the superconducting state is indeed unique and shows consistency with the experimental results. The trivial *dV/dI* spectra of graphene without any sub-peaks would result in one-step increase in resistance when driving superconducting to normal state. Such usual behavior indicates that Cooper pairs are broken into quasiparticles in an abrupt manner, different from multiple-step process occurring in the superconducting WTe$_2$ thin flake. Combining all the evidence above, we believe that the nontrivial signature in the *dV/dI* spectra of WTe$_2$ is unique and roots in its intrinsic band structure. Note that the sub-gap features in WTe$_2$ may occur in another type-II Weyl semimetal MoTe$_2$[55], thus this should stimulate further exploration of the properties of type-II Weyl semimetals in superconducting phase.

In summary, we achieve superconductivity for the first time in type-II Weyl semimetal $WTe_2$ by proximity effect based on a $WTe_2$/$NbSe_2$ heterostructure. The superconducting transition occurs at 6.4 K in $WTe_2$, close to $T_c$ of $NbSe_2$. Most importantly, we discover anomalous oscillations of *dV/dI* spectra in superconducting $WTe_2$. Furthermore, our theoretical calculations confirm that such peculiar feature in the *dV/dI* spectra in superconducting $WTe_2$ results from its nontrivial DOS spectrum. This work suggests that superconductivity in type-II Weyl semimetals exhibits distinctive feature from conventional superconductors and will stimulate further interest in exploration of topological superconductivity.

We grew the high-quality $NbSe_2$ single crystals with a standard chemical vapor transport method. High purity powders of elements Nb and Se are stoichiometric mixed and thoroughly grounded. Then the mixture was sealed under vacuum in a quartz tube with iodine (5mg cm$^{-3}$). The tube was placed into a two-zone horizontal furnace with the hot end of 780 °C and the cold end of 700 °C for two weeks. High-quality plate-like $NbSe_2$ single crystals are obtained.

Single crystals of $WTe_2$ were grown by a high temperature self-flux method. High-purity W powders (99.9%) and Te pieces (99.999%) were put in alumina crucibles with a molar ratio of 1:30 in a glove box filled with pure argon then sealed in quartz tubes under high vacuum. The tubes were heated to 1373 K and maintained for 10 hours. Then the furnace was slowly cooled down to 923 K with a rate of 2 K/h followed by separating the Te flux in a centrifuge at 923 K.

We mechanically exfoliated h-BN onto a highly doped Si wafer covered by a 300-nm-thick $SiO_2$ layer. The bottom electrodes (5 nm Ti/40nm Au) were patterned on h-BN flake using standard electron beam lithography method and deposited by standard electron beam evaporation. The $WTe_2$, $NbSe_2$ and graphene thin flakes were mechanically exfoliated from single crystals onto the PDMS (Polydimethy lsiloxane) substrate in a glove box filled with an inert atmosphere, followed by dry transferring on the top of bottom electrodes layer by layer. Finally, a Boron Nitride thin flake was transformed on to $WTe_2/NbSe_2$ heterostructure to avoid the device degradation in the air. All the transfer processes mentioned above were finished in the glove box to achieve the best quality of the device. Bruker Multimode 8 atomic force microscopy was used to identify the thickness of $WTe_2$, $NbSe_2$ and graphene thin flakes.

The devices were measured in an Oxford cryostat with a base temperature of about 1.6 K. The resistance signals were collected using a low-frequency Lock-in amplifier. The differential resistance signals were collected by applying an AC excitation current of 1 nA added on top of a variable DC bias current.

## ASSOCIATED CONTENT

**Supporting information**

The Supporting information is available free of charge on the ACS Publications website at http://pubs.acs.org.

Superconductivity in $WTe_2/NbSe_2$ heterostructure with a trench in $NbSe_2$ layer; Low-temperature electrical transport properties of $NbSe_2$ under different temperatures and magnetic fields; Mobility for $WTe_2$ encapsulated by BN flakes; Critical current and observation of sub-gap anomaly in other $WTe_2$ samples; Theoretical calculation methods for DOS of s-wave superconductor ($NbSe_2$) and Weyl semimetals (WSM); DOS spectra of $NbSe_2$ and type-I Weyl semimetal; *I-V* characteristics of graphene thin flake under various temperatures and magnetic fields.

## AUTHOR INFORMATION

**Corresponding Authors**

*Tel: +86-025-83621497. Fax: +86-025-83621497. E-mail: miao@nju.edu.cn (F. M.); qhwang@nju.edu.cn (Q. H. W.); sjliang@nju.edu.cn (S. J. L.).

**Notes**

The authors declare no competing financial interests.

**Acknowledgements**

This work was supported in part by the National Key Basic Research Program of China (2015CB921600), the National Natural Science Foundation of China (61625402, 61574076, 11774399, 11474330, 11574134, 11534010), the Key Research Program of Frontier Sciences CAS (Grant No. QYZDY-SSW-SLH021), the National Key Research and Development Program of China (2016YFA0300401, 2017YFA0302901, 2016YFA0300604), Fundamental Research Funds for the Central Universities (020414380093, 020414380084), the Collaborative Innovation Center of Advanced Microstructures and Natural Science Foundation of Jiangsu Province (BK20180330). K.W. and T.T. acknowledge support from the Elemental Strategy Initiative conducted by the MEXT, Japan and the CREST (JPMJCR15F3), JST.

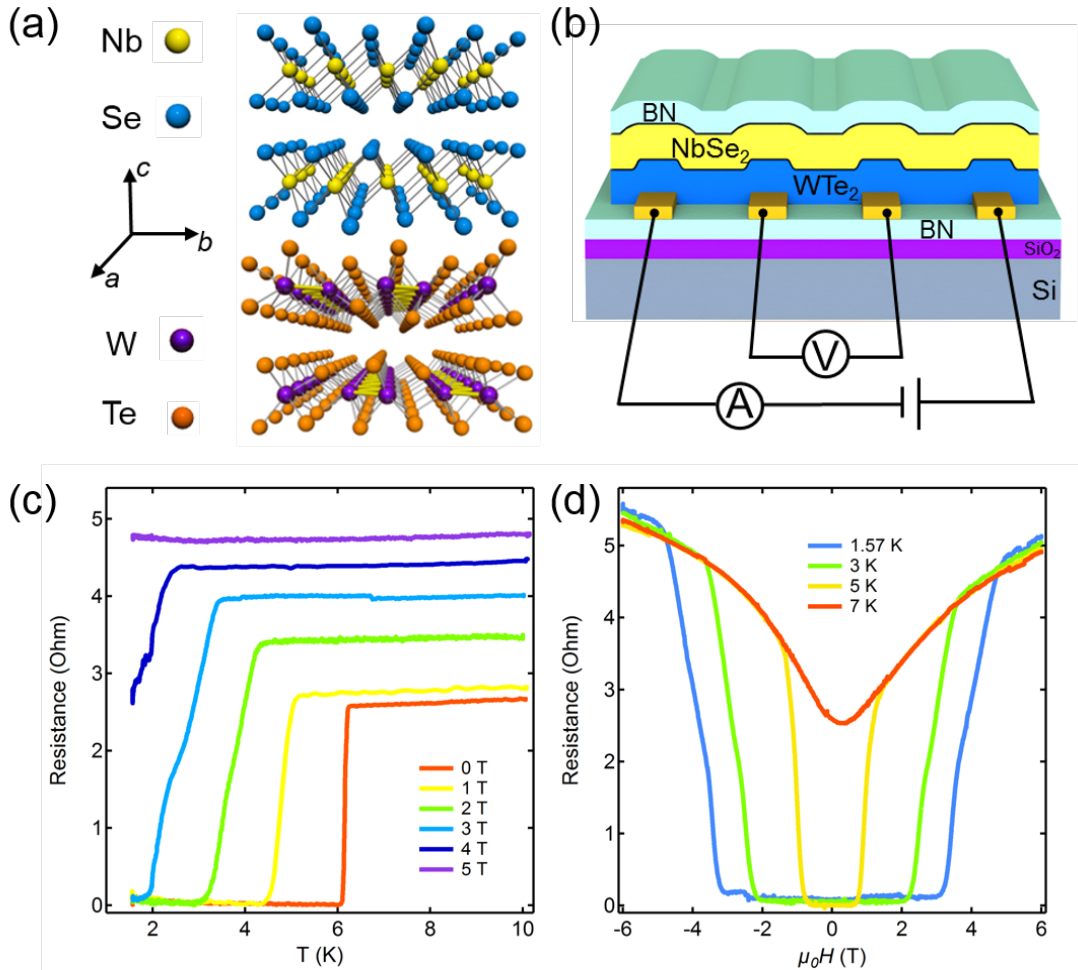

**Figure 1.** WTe$_2$/NbSe$_2$ device and basic superconducting characterizations of WTe$_2$. (a) Crystal structure of WTe$_2$ and NbSe$_2$. (b) Schematic structure and measurement setup of a four-probe WTe$_2$/NbSe$_2$ device. (c) *R-T* curves of WTe$_2$ channel at different applied perpendicular magnetic fields. (d) *R-$\mu_0H$* curves of WTe$_2$ channel measured at various temperatures. The magnetoresistance above T$_c$ acts like weak anti-localization of WTe$_2$.

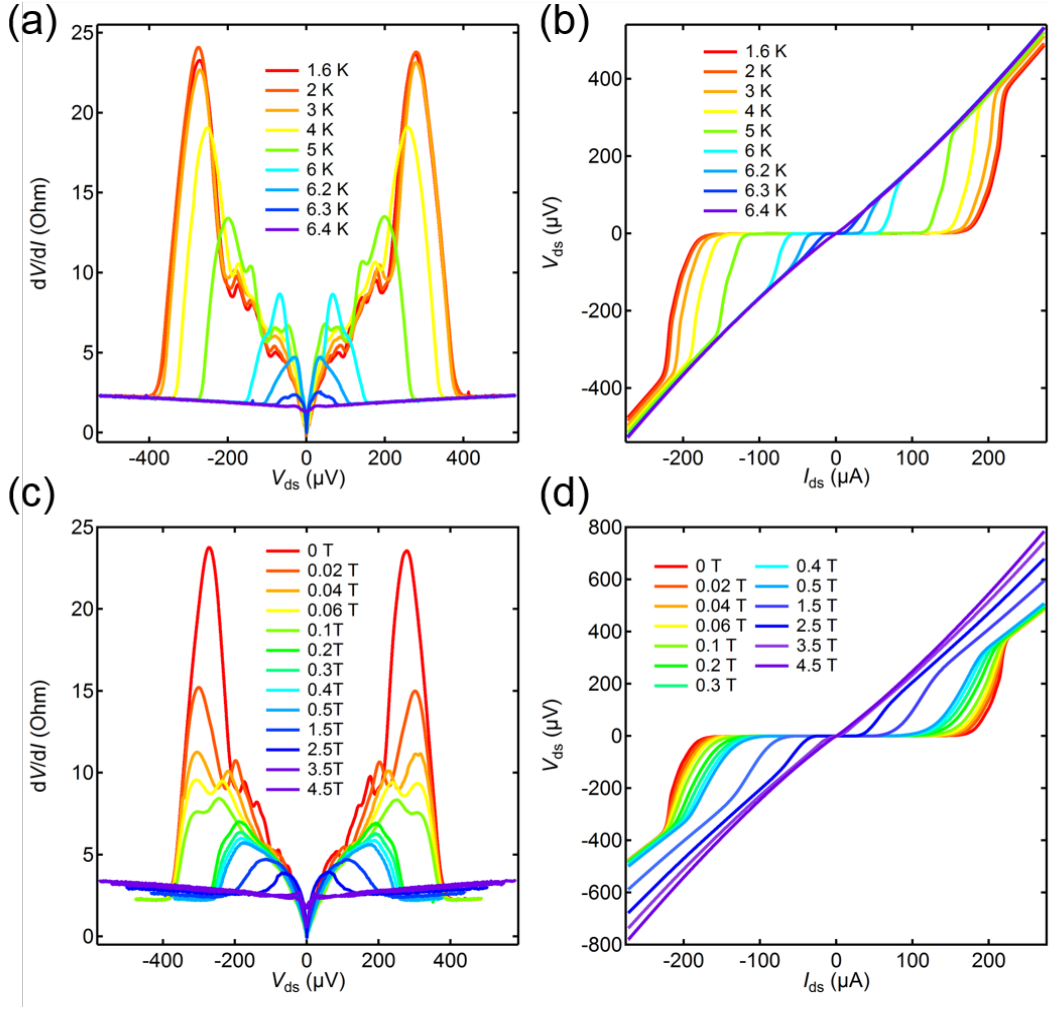

**Figure 2.** Differential resistance spectra and Current-Voltage (I-V) curves of $WTe_2$. (a) Differential resistance spectra of $WTe_2$ measured at various temperatures. Several smaller peaks emerge inside of two main peaks. (b) *I-V* curves of $WTe_2$ at different temperatures ranging from 1.6 K to 6.4 K without applying magnetic field. (c) Differential resistance spectra of $WTe_2$ measured at various applied magnetic fields at 1.6 K. (d) *I-V* curves of $WTe_2$ at different applied magnetic fields ranging from 0 T to 4.5 T.

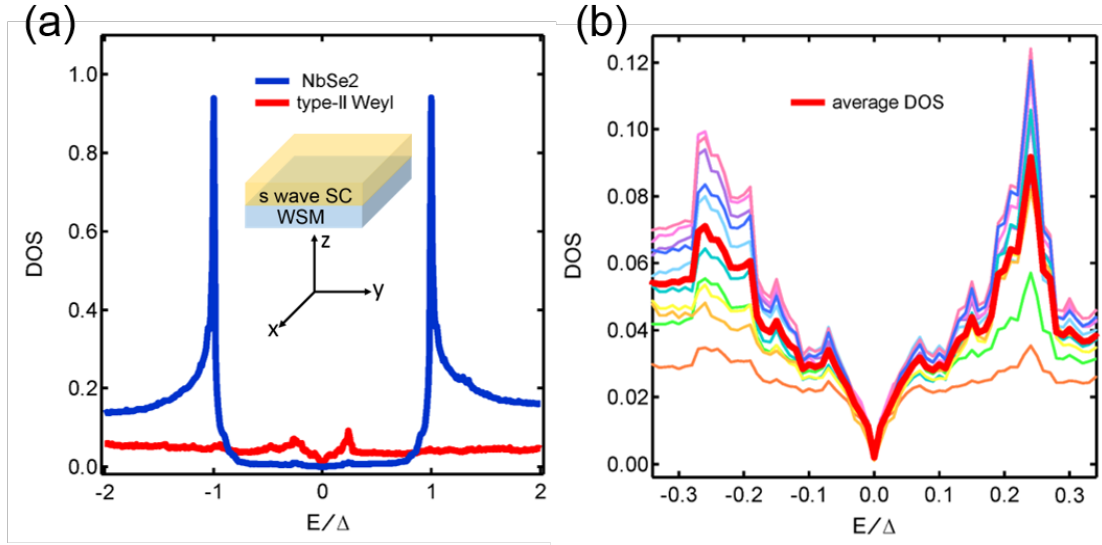

**Figure 3.** Theoretical calculations of density of states in superconducting WTe$_2$/NbSe$_2$ heterostructure. (a) Density of states (DOS) spectra (in arb. unit) versus energy E (normalized by the maximal gap Δ) of type-II Weyl semimetal (WTe$_2$) and s-wave superconductor (NbSe$_2$) in superconducting state. E/Δ=1 is the superconducting gap we included in the calculations. Inset shows schematic drawing of theoretical calculation model. (b) Enlarged DOS spectra of WTe$_2$ in (a) (red line). Thin curves represent DOS of different layers in WTe$_2$.

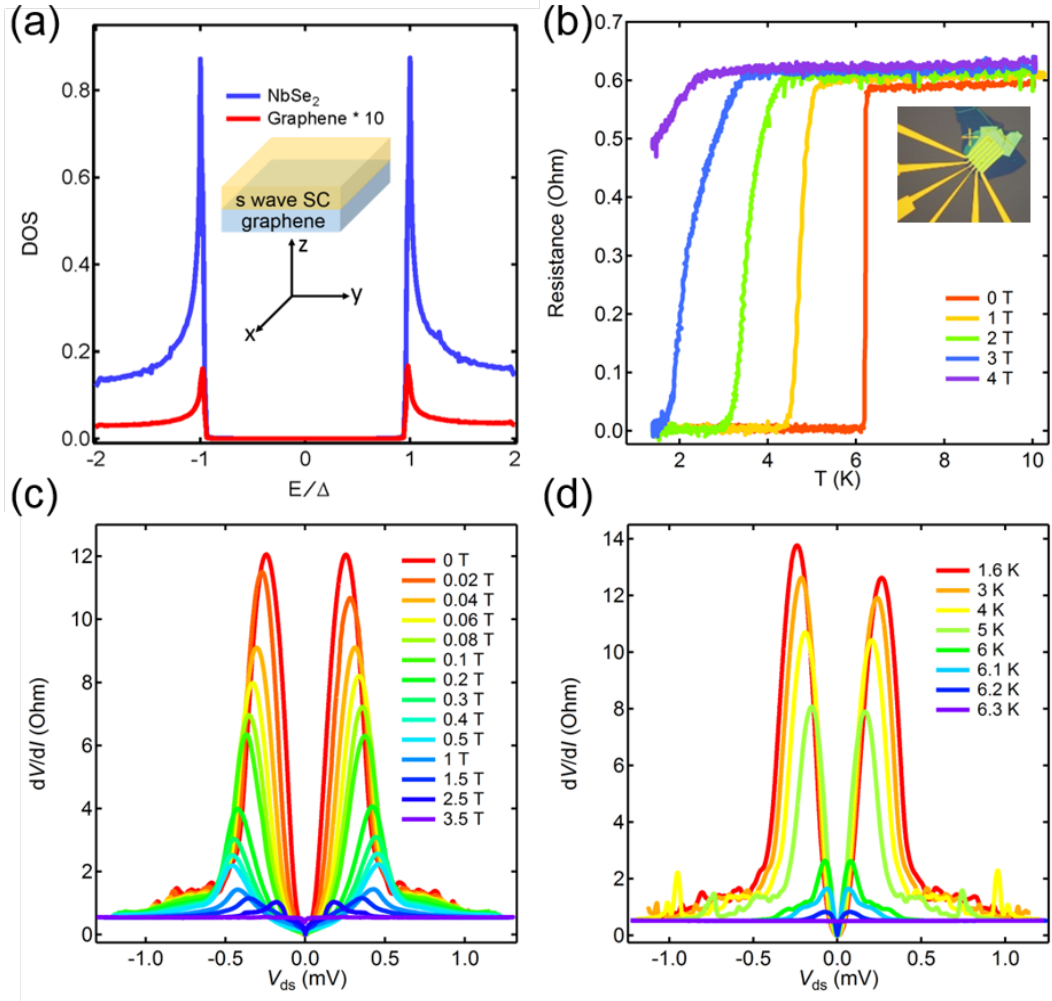

**Figure 4.** Theoretical calculations and experimental data of graphene thin flake. (a) DOS of graphene thin flake and NbSe$_2$ in superconducting state, showing a trivial (DOS) signature. Inset: schematic drawing of theoretical calculation model. (b) *R-T* characteristics of graphene thin flake (around 10 nm) at various magnetic fields ranging from 0 T to 4 T. Inset: optical image of graphene/NbSe$_2$ heterostructure device. (c) Differential resistance spectra of graphene thin flake at different temperatures. The straight line at 6.4 K indicates that the superconductivity is completely suppressed. (d) Differential resistance spectra of graphene thin flake at different applied magnetic fields.

For TOC graphic only

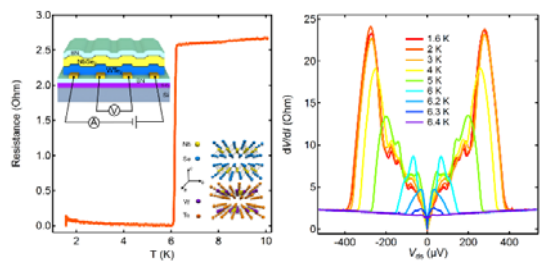